\documentclass[%
	 reprint,
	superscriptaddress,
	preprintnumbers,
	nofootinbib,
	 amsmath,amssymb,
	 aps,
	floatfix,
	multicol,
	]{revtex4-1}
	
\usepackage[italicdiff]{physics}
\usepackage{graphicx,color,hyperref}
\usepackage{dcolumn}% Align table columns on decimal point
\usepackage{bm}% bold math

\begin{document}

\preprint{KUNS-2888, OU-HET 1101}

\title{
	Deriving dilaton potential in improved holographic QCD from meson spectrum
}
% Force line breaks with \\

\author{Koji Hashimoto}%
\email{koji@scphys.kyoto-u.ac.jp}
\affiliation{%
Department of Physics, Kyoto University, Kyoto 606-8502, Japan
}
\author{Keisuke Ohashi}%
\email{keisuke084@gmail.com}
\affiliation{%
Research and Education Center for Natural Sciences, Keio University,
%Hiyoshi 4-1-1, 
Kanagawa 223-8521, Japan
}
\author{Takayuki Sumimoto}%
\email{t\_sumimoto@het.phys.sci.osaka-u.ac.jp}
\affiliation{%
Department of Physics, Osaka University, Toyonaka, Osaka 560-0043, Japan
}

\begin{abstract}
	We derive  an explicit form of the dilaton potential in improved holographic QCD (IHQCD) 
	from the  experimental data of the $\rho$ meson spectrum. 
	 For this purpose we make use of the emergent bulk geometry obtained by deep learning from the hadronic data  in the work by Akutagawa {\it et.al}
	\cite{Akutagawa:2020yeo}.
	% \url{https://arxiv.org/abs/2005.02636}
	 Requiring that the  geometry is a solution of  an IHQCD derives the corresponding dilaton potential backwards.
	This determines the bulk action in a data-driven way, which enables us at the same time to ensure that the deep learning proposal is a consistent gravity.
	Furthermore, we find that the resulting potential satisfies the requirements normally imposed in IHQCD, and that the holographic Wilson loop for the derived model exhibits quark confinement.

	\vspace*{10mm}
\end{abstract}

\maketitle

%\tableofcontents

%%%%%%%%%%%%%%%%%%%%%%%%%%%%%%%%%%%%%%%%%%%%%%

\section{Introduction}
\label{sec:intro}
	The advent of the AdS/CFT correspondence \cite{Maldacena:1997re,Gubser:1998bc,Witten:1998qj} has brought new aspects to various theories, QCD being one of them. Methodologies to describe QCD phenomena with models inspired by the AdS/CFT correspondence have been studied. Among them, the AdS/QCD model is a phenomenological modification of the AdS spacetime making a conformal symmetry of the bulk broken to describe the physics of hadrons and QCD.

	The simplest example is the hard wall model in which a scale is introduced by a cutoff wall in the radial location of the pure AdS spacetime \cite{Erlich:2005qh}. In this model, the meson spectrum is reproduced with some accuracy, and the chiral symmetry breaking is described by a hand-added boundary condition at the wall.  It opened the possibility of describing QCD low energy physics by building the IR region of the spacetime in the holographic models.

	Immediately after \cite{Erlich:2005qh}, in the soft wall model \cite{Karch:2006pv}, a non-constant dilaton was introduced. In this model the conformality is broken due to the scale introduced in the dilaton field profile. The properties of the model are %fixed 
%determined by the engineering of 
	engineered by the functional form of the dilaton field. The authors of \cite{Karch:2006pv} showed that a specifically chosen dilaton profile %quadratic function of the radial coordinate
	reproduces the linear confinement with respect to the meson spectrum.

	Although these models are based on the holographic principle, they treat the gravitational fields only as background fields, which is contrary to
%  the opposite approach to
	the string theory-based holographic QCD \cite{Sakai:2004cn,Kruczenski:2003uq}. As a hybrid of these two approaches, IHQCD was proposed \cite{Gursoy:2007cb,Gursoy:2007er}, which is based on the five-dimensional Einstein-dilaton system with a non-trivial dilaton potential.
	Beyond the previous AdS/QCD models, this approach enables one to treat the physics at various temperatures comprehensively, %and it is even possible to analyze 
	even including the QCD phase transition \cite{Gursoy:2008bu,Gursoy:2008za}.
	Furthermore, since the action of the bulk theory is specified, one can calculate various thermodynamic quantities at once from the bulk partition function, and a good agreement with the lattice QCD results was reported \cite{Gursoy:2007cb,Gursoy:2007er}.
	In the minimal Einstein-dilaton system, only the dilaton potential determines the model, and fixing it completely determines %the described QCD dynamics.
	the model prediction for QCD observables.

	Therefore, in order to arrive at a holographic model describing QCD, it is necessary to determine the form of the appropriate dilaton potential. For example, in the IHQCD there is a method of determining the IR and UV asymptotic forms of the potential so that they satisfy the necessary conditions as a dual of QCD, and then interpolating them. While this is a direct method of determining the potential, there exists a different method %it is also possible 
	to determine the potential by solving Einstein's equation backwards with a given metric, dilaton or a beta function for the `t Hooft coupling \cite{Alanen:2009xs,Ballon-Bayona:2017sxa}.
	In spite of these efforts, the dilaton potential %  In any case, it 
	has been not completely fixed as a fate of the bottom-up approach, and a more systematic method is needed to determine the QCD dual.

	For this purpose, we present a method to derive the dilaton potential from the QCD observable data. We use the profiles of the background gravitational fields determined so that the bulk probe field fluctuation can reproduce the QCD observable data. The profiles are regarded as a solution of %substituted into 
	the Einstein equations.
	This requirement derives the dilaton potential backwards.
	As for the gravitational field profiles determined from the data, we can use the results using machine learning \cite{Akutagawa:2020yeo}.
	% Since the gravitational field is a degree of freedom that should be determined by the Einstein equation, we directly calculate the dilaton potential, which is an indefinite function of the model.
	It is important that %it is possible to determine the potential in a data-driven manner.
	the dilaton potential can be derived in a data-driven manner.

	In the machine learning calculation \cite{Akutagawa:2020yeo}, the zero-temperature holographic model is characterized by two background fields, the metric and dilaton fields, and their properties are determined by the consistency with the meson spectrum.
	% calculate their configuration.
	The line element is parametrized as
	\\*
	\begin{align} \label{eq:string frame metric}
		ds^2 = e^{2A(z)} \qty( dz^2 + \eta_{\mu\nu}dx^\mu dx^\nu ).
	\end{align}
	\\* 
	In this soft wall model, 
%  In such a model, represented by the soft wall model, 
	a useful combination of $A(z)$ and the dilaton field $\Phi(z)$ is 
	\\*
	\begin{align} \label{eq:B}
		B(z) \equiv \Phi(z)- A(z).
	\end{align}
	\\*
	The vector meson spectrum is calculated by solving the equation of motion of a bulk probe massless vector field.
	Therefore, determining the optimized model for given meson spectra is a kind of inverse problems, and thus algorithms based on neural networks are useful with spectral data as their input.
	Once the vector field equation is written down in a discrete form, one can translate the components of the equation into the components of the neural network, especially the background fields into the network weights \cite{Hashimoto:2018ftp}. Here the weight is a quantity that can be optimized in training the neural network.
	In the present case, because the vector field equation includes the background fields only in the form of $B'(z)$, this $B'(z)$ is identified to weights. As a result
	of the machine learning with the $\rho$ meson spectrum as the training data, %we were able to obtain 
	the explicit profile of $B'(z)$ is determined \cite{Akutagawa:2020yeo}.

	The purpose of this paper is to derive a dilaton potential by this $B'(z)$ profile.
	The method is as follows. First, we write down the equations of motion including the dilaton potential. %using the optimized $B'$ for the meson.
	Since we are assuming a zero-temperature system as the spacetime, the gravitational field has only one degree of freedom. Together with the dilaton's degree of freedom, the Einstein-dilaton system has two degrees of freedom, so we can obtain two independent equations of motion. Since the dilaton potential is undecided, we first need to eliminate it from these two equations.
	%, but we cannot solve the equations as they are. Therefore, 
	Then we use the $B'(z)$ profile to reduce that equation to that of the dilaton only, and solve it to obtain $\Phi(z)$. We substitute it into the equation including the potential, and finally derive the dilaton potential $V(\Phi)$.
	% Thus, we end up with the derivation of the dilaton potential from the given $B'(z)$.

%  When we write down the equation, we know that we will also need the information of B, but we will treat the integration constant of B' as a parameter. 
	
	At the end of this paper, we investigate the physics described by the derived Einstein-dilaton system in the following two perspectives: the asymptotic form of the obtained potential and the holographic Wilson loop that can be calculated from the derived model.

%%%%%%%%%%%%%%%%%%%%%%%%%%%%%%%%%%%%%%%%%%%%%%

\section{Derivation of dilaton potential}
\label{sec:zeroT}

	We are going to study an Einstein-dilaton system 
	%effective theory of gravity sector 
	in the five dimensional spacetime with an undetermined dilaton potential.
	In this section, we determine the dilaton potential $V(\Phi)$ so that the $B'(z)$ profile given by the data of hadron spectra is a solution of the Einstein equations. In addition, we would like to investigate the IR asymptotic behavior of the dilaton potential in comparison to the standard IHQCD potential.
	We also find that the resultant metric function indicates that the QCD confinement can be described by the derived holographic model.

	In a minimal approach of IHQCD, one starts with a five dimensional Einstein-Hilbert action accompanying a dilaton sector.
	This includes a five dimensional metric $g_{MN}$, which is dual to the stress tensor operator in QCD, and the dilaton field $\Phi$, which is associated with the gluon composite operator.
	The Einstein frame action %up to overall constant 
	is
	\\*
	\begin{align} \label{eq:action}
		S & = c  \int d^5x \sqrt{g} \qty\Big(R - \frac{4}{3}g^{MN}\partial_M\Phi\partial_N\Phi + V(\Phi)).
	\end{align}
	\\*
	The overall constant $c$ is the gravitational constant which is irrelevant in our work. 
	% Here $V(\Phi)$ is the effective potential coming from, in principal, string theory on curved spacetime.
	In the standard holographic model building, the dilaton potential $V(\Phi)$ is provided
%  People usually determine it 
	so that the resultant system or the geometry reproduces known properties in QCD. 
	Specifically, the behavior of the potential at a large $\Phi$ region is known to be important to reproduce the Regge behavior of the glueball spectrum.
	According to \cite{Gursoy:2007cb,Gursoy:2007er}, when $\Phi$ is large $V(\Phi)$ should take the form up to an overall factor
	\\*
	\begin{align} \label{eq:IR behavior}
		& V(\Phi) \sim e^{2Q\Phi} \Phi^P 
		%\\
		%& \qq{with} P=1/2, ~ Q=2/3.
	\end{align}
	\\*
	with $P=1/2$ and $Q=2/3$.

	We assume a five dimensional asymptotically AdS spacetime as in the standard AdS/QCD approach.
	% the bulk geometry computed in \cite{Akutagawa:2020yeo}. Let us review the setup and the result.
	Since we are going to explore the system %whose the given \eqref{eq:B} is the solution, 
	whose solution is a given explicit function $B(z)$ in \eqref{eq:B}, 
	we should use the same gauge as that used in \cite{Akutagawa:2020yeo}:
	\\*
	\begin{align} \label{eq: string frame}
		ds^2 = e^{2A(z)-\frac{4}{3}\Phi(z)} \qty( dz^2 + \eta_{\mu\nu}dx^\mu dx^\nu ).
	\end{align}
	\\*
	Note that this is the Einstein frame metric, equal to string frame metric \eqref{eq: string frame} times $e^{-4\Phi/3}$.
	The coordinate $z$ is the direction transverse to the spacetime of the boundary, and everything is made dimensionless in the unit of the AdS radius $L$.
	As for the warp factor $A(z)$, we shall impose the following two conditions according to \cite{Akutagawa:2020yeo}. 
	First, the UV cutoff is put at $z=z_c\equiv 0.2$ then the bulk region is restricted to $z \ge z_c$ in the following calculations.
	Second, as the asymptotic AdS condition, we shall set $A(z) = - \log z$ at $z=z_c$.

	In the gauge \eqref{eq: string frame}, the Einstein equations for \eqref{eq:action} consist of two independent equations:
	\\*
	\begin{align} \label{eq:EE}
		\begin{cases}
			6A'' + 6A'^2 - 8A'\Phi' - 4\Phi'' + 4\Phi'^2 - e^{2A-\frac{4}{3}\Phi} V = 0 \\
			12A'^2 - 16A'\Phi' + 4\Phi'^2 - e^{2A-\frac{4}{3}\Phi} V = 0.
		\end{cases}
	\end{align}
	\\*
	Here we use primes to denote a derivative with respect to $z$.
	We should notice that one can derive one more equation of motion by variation of the action with respect to $\Phi$, though it turns out to be unnecessary because it is not independent.
	Note that we have only two degrees of freedom in the present model due to the metric assumption of \eqref{eq: string frame}.

	Our goal is to find a dilaton potential $V$ as a function of $\Phi$.
	For some given configuration of $A$ or $\Phi$, one might determine the dilaton potential as a functional of $\Phi$.
	In our case, since we have the numerical information about %$A$ or $\Phi$, 
	$B'(z)$,
	we compute the potential $V(z)$ numerically and investigate the dependence on $\Phi$ by drawing $V$ against $\Phi$ with $z$ treated as a parameter.
	
	% In \cite{Akutagawa:2020yeo}, supposed the metric and the dilaton field were background, the values of $B'(z)\equiv\Phi'(z)-A'(z)$ on discrete points were estimated. 
		% \footnote{To tell the truth we computed bulk gravity by taking advantage of supervised learning in \cite{Akutagawa:2020yeo}, in which training data was the spectra of $\rho,~a_2$ meson. 
		% Having focused on the difference of spin, we have obtained $A(z)$ and $\Phi(z)$ individually.
		% But we decide not to adopt the result for $a_2$ meson because it was highly depending on the artificial regularization.}
	% Here what we want to use 
	More precisely the numerical information is the value of $B'(z)\equiv\Phi'(z)-A'(z)$ at the each discrete point on the $z$ axis. In order to proceed, we fit those values to a smooth function. The simplest choice is a polynomial. The fifth order polynomial fits those values well, % and we call the best-fit model $B'(z)$. That is
	% More precisely, the result was actually
	\\*
	\begin{align} \label{eq:Bprime}
		B'(z) = \sum_{p=0}^5 a_p z^p
	\end{align}
	\\*
	with the coefficients shown in TAB.~\ref{tab:coefficients}.

	\begin{table}[tb]
		\begin{tabular}{c|c|c|c|c|c|c}
			\hline
			$p$ & 0 & 1 & 2 & 3 & 4 & 5 \\
			\hline \hline
			$a_p$ & 8.702 & -22.928 & 24.504 & -9.726 & 1.670 & -0.102 \\
			\hline
		\end{tabular}
		\caption{List of coefficients for the polynomials in \eqref{eq:Bprime}}
		\label{tab:coefficients}
	\end{table}

	Eliminating the unknown potential $V$ from \eqref{eq:EE} and using \eqref{eq:B}, we obtain a differential equation for the dilaton field:
	\\*
	\begin{align} \label{eq:reduced EE}
		\Phi'' = -\Phi'^2 -2B'\Phi' +3B'^2 +3B''.
	\end{align}
	\\*
	Since this is a non-linear equation, we carry out a numerical method for solving this equation. Here we simply integrate the equation step by step from the UV cutoff. The initial condition for $\Phi'$ should be $\Phi'(z_c)=0$ because it was employed in \cite{Akutagawa:2020yeo}. On the other hand, the initial condition for $\Phi$ is arbitrary. So we name $\Phi_c \equiv\Phi(z_c) $ and treat it as a free parameter below.\footnote{
	In the standard IHQCD the dilaton goes to minus infinity $(\lambda = e^\Phi\to0)$ at the UV boundary while we introduce the UV cutoff at $z = z_c$ so the precise UV behavior is beyond the scope of this paper.
	% {
	% % Our dilaton goes to a constant at the UV boundary, while in 
	% the standard IHQCD
	% the dilaton goes to minus infinity ($\lambda = e^\Phi \to 0$). However, this difference may not be so significant as we impose the UV cutoff at $z=z_c$.
	% }
	}
	
	% We evaluate $\Phi$ on some discrete points and fit those values to the fifth order polynomial again. Then, we get the dilaton field
	% \begin{align} \label{eq:Phi}
	%     \Phi(z) = \sum_{p=0}^5 b_p z^p
	% \end{align}
	% with coefficients $b_i$ shown in TAB.~\ref{tab:coefficients}.

	We would like to mention how we calculate $B=\Phi-A$. Obviously $B$ can be obtained by integrating $B'(z)$, but a subtle point is its integration constant.
	% We may take it so that $B(0.2) = \Phi(0.2)-A(0.2)$, where $A(0.2)$ is fixed to $-\log(0.2)$ by the asymptotic AdS condition. So its expression is
	To ensure $B(z_c) = \Phi(z_c)-A(z_c)$, we define $B(z)$ as
	\begin{align} \label{eq:B as integrate}
			B(z) \equiv \int^z_{z_c} B'(\tilde{z}) d\tilde{z} + \Phi_c + \log(z_c),
	\end{align}
	where $A(z_c)$ is set to $-\log(z_c)$ by the asymptotic AdS condition.

\begin{figure}[tb]
	\centering
	\includegraphics[width=0.95\linewidth]{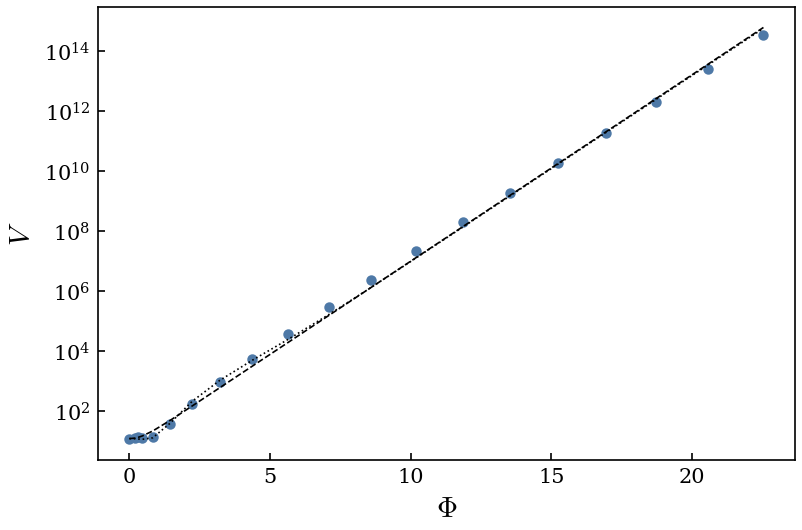}
	\caption{
		Plot of $V(\Phi)$ in a log scale.
		The dashed line and dotted line represents \eqref{eq:fit V to cosh 1} and \eqref{eq:fit V to cosh 2} individually.
		One can see that the exponential dependence on $\Phi$ is dominant at large $\Phi$. 
		% This is consistent %behavior to be valid against to with the condition \eqref{eq:IR behavior}.
	}
	\label{fig:potential}
\end{figure}

	Finally we evaluate the dilaton potential.
	From \eqref{eq:EE} and $B'(z)=\Phi'(z)-A'(z)$, the potential can be written as follows:
	\\*
	\begin{align} \label{eq:potential}
		% V = 4\exp\qty(-\frac{2}{3}\Phi+2B) \qty(-2B'\Phi' + 3B'^2).
		V = 4 e^{-\frac{2}{3}\Phi+2B} \qty(-2B'\Phi' + 3B'^2).
	\end{align}
	\\*
	Combining this with \eqref{eq:Bprime} and $\Phi(z)$ obtained from \eqref{eq:reduced EE} leads to find $V(z)$.
	Then, by plotting $V$ against $\Phi$ with $z$ being a parameter, we finally obtain $V(\Phi)$. Its plot is shown in FIG.~\ref{fig:potential}.

	Potential fitting
	For the comparison between our obtained potential and various AdS/QCD models,
	here we provide a smooth analytic 
	% To obtain more useful result, it is better to proceed to find a smooth function 
	fitting the whole obtained plot.
	To make the fitting independent of $\Phi_c$, we use
	\\*
	\begin{align} \label{eq:}
		& \tilde{V} \equiv  e^{-4\Phi_c / 3} V, \quad  
		X \equiv \Phi - \Phi_c.
	\end{align}
	\\*
	Obviously, $\tilde{V}(X) = V(\Phi)$ for $\Phi_c=0$ which is used in FIG.~\ref{fig:potential}.
	Since we see that the growth of the obtained potential is mostly linear in the log plot, we can approximate its behavior with an exponential function.
	One candidate which can simultaneously fit also the part close to \text{$X=0$} is
	\\*
	\begin{align} \label{eq:fit V to cosh 1}
		\tilde{V} = 12 \cosh \qty(\gamma X).
	\end{align}
	\\*
	and the best-fit value of the parameter is $\gamma=1.433$.
	Interestingly, such a dilaton potential was used in \cite{Gubser:2008ny}
	which provide the prediction to QCD quantities from an Einstein-dilaton system.
	\footnote{
	As pointed out in \cite{Gubser:2008ny}, when one chooses the single hyperbolic cosine potential and considers the propagation of the dilaton field, $\gamma$ is bounded by the Breitenlohner-Freedman (BF) bound \cite{Breitenlohner:1982bm,Breitenlohner:1982jf} because the dilaton mass depends only on $\gamma$.
	Unfortunately, for our best-fit value of $\gamma$ in the case of \eqref{eq:fit V to cosh 1}, the dilaton mass breaks the bound.
	This motivates us to incorporate additional terms to avoid this problem, and one of the possible modifications is
	\begin{align} \label{eq:fit V to cosh 2}
		\tilde{V} = 12 \cosh \qty(1.430X) -16.778 ~ X^2 + 5.943 ~ X^4.
	\end{align}
	This also fits our numerical data well, while satisfies the BF bound.
	}

\begin{figure}[tb]
		\centering
		\includegraphics[width=0.95\linewidth]{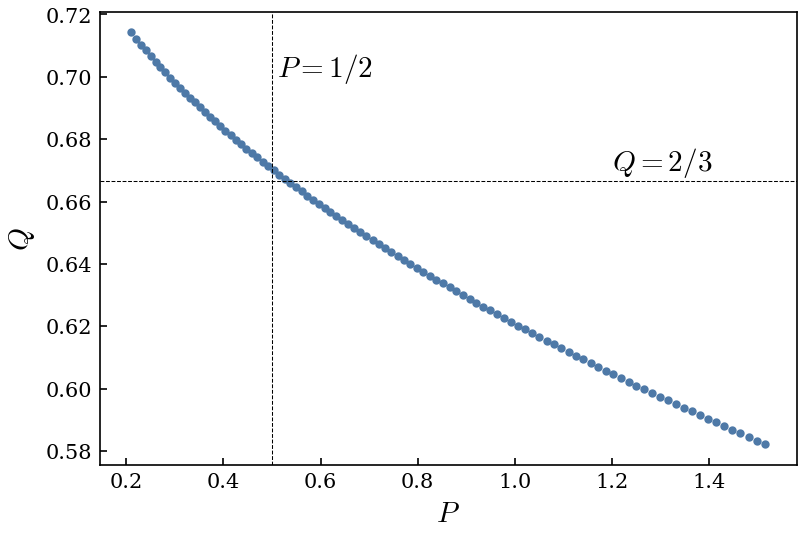}
		\caption{
			The values of $(P,Q)$. We use 200 initial values which divide [-2,0] into equal intervals. Each blue point corresponds to the result of regression with each of the initial values. As the initial value grows, $P$ tends to increase and $Q$ decreases.
		}
		\label{fig:PQ}
	\end{figure}

Let us compare the derived potential with the one used in the IHQCD, 
	%In order to verify the satisfaction of the condition 
	\eqref{eq:IR behavior}.
We shall perform a non-linear regression of $V(\Phi)$ with \eqref{eq:IR behavior} and evaluate the values of $(P,Q)$.
	The subtle point is that $V(\Phi)$ has an ambiguity coming from $\Phi_c$. When $\Phi_c$ changes $V(\Phi)$ varies by some overall constant.
	% But this does not affect our task of checking whether $V(\Phi)$ behaves in the way of \eqref{eq:IR behavior}.
	Due to this ambiguity, 
	% when one has started with a different $\Phi_c$, 
	the values of $(P,Q)$ depend on $\Phi_c$. So we calculate $(P,Q)$ for various initial values.
	The result when fitting $V$ to \eqref{eq:IR behavior} in the range $1.4\le z\le 2$ is shown in FIG.~\ref{fig:PQ}.\footnote{Having chosen this region is because 1) $\Phi$ needs to be large enough and 2) the deep learning result is reliable.}
	% \footnote{Need more explanation.}
	% But, this is just one example and change of the $z$ range makes the distribution of $P,Q$ different(See FIG. \ref{fig:PQs}).
	There are 200 points representing the values $(P,Q)$ with 200 independent $\Phi_c$'s. 
	This plot shows there exists a pair of $(P,Q)$ which almost equals $(1/2,2/3)$.
	% one in the condition \eqref{eq:IR behavior}
	% We can find the fit result matching to 
	The point is located at $P = 0.50, ~ Q = 0.67$ and the corresponding value of $\Phi_c$ is -1.43. 

	%  One might be skeptical that $P$ and $Q$ depend on $\Phi_c$ in spite of trivial dependence as shown in \eqref{eq:potential}. This happens because our fitting procedure only see a restricted region. The fit model \eqref{eq:IR behavior} includes power of $\Phi$.   
	%  Then when the region of $\Phi$ for fit shifts by changing $\Phi_c$, the best fit parameters except for overall constant also vary.
	%  This is of course artificial error but we cannot decrease it at current stage. 
	%  The situation may be better or worse if the bulk gravity be reconstructed beyond z=4.
	%  In any case we can see the possibility that the obtained dilaton potential satisfies the requirement \eqref{eq:IR behavior}.
	%   \\*
	%   \begin{align} \label{eq:PQ value}
	%     P = 0.50, ~ Q = 0.67.
	%   \end{align}
	%   \\*
	%   This result insists that the predicted potential from the geometry which machine learned can be suitable for the dilaton potential of IHQCD.
	%   Therefore we can conclude it is possible for the machine-learned metric to be interpreted as a solution of the minimal Einstein-Dilaton system, not just an artificial.

	%   So we can conclude we have found the dilaton potential from machine learned spacetime and it has IR behavior acceptable in the context of IHQCD.

\begin{figure}[tb] 
		\centering
		\includegraphics[width=0.95\linewidth]{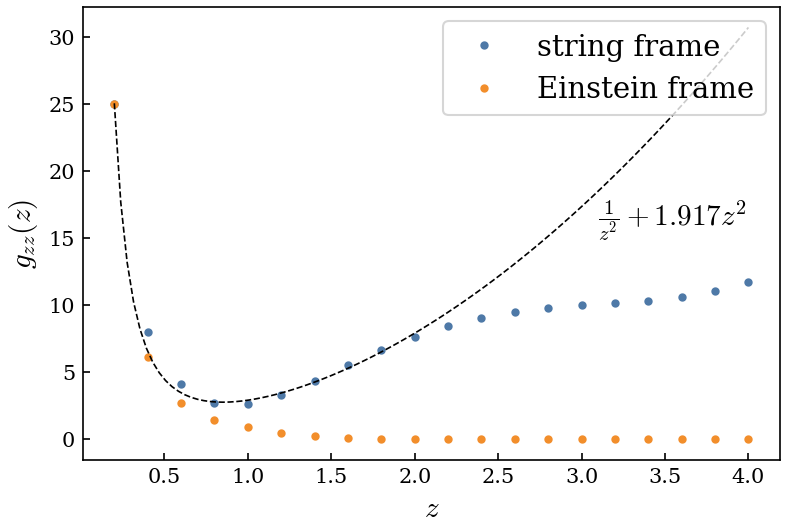}
		\caption{
			The metric obtained by solving \eqref{eq:EE} for our given $B(z)$. The string frame metric means $\exp(2A(z))$ and the Einstein frame metric corresponds to $\exp(2A(z)-\frac{4}{3}\Phi(z))$.
			In this figure $\Phi_c$ is set to 0 just to make it easier to compare them.
			% as one can see from the fact that $g_{zz}(0.2)$ for both curves matches.
			The dashed line represents \eqref{eq:fit gtt}.
		}
		\label{fig:metric}
	\end{figure}

	Before ending this section, we study in detail the metric of the obtained model. In the way to calculate the dilaton potential, we have solved Einstein equation \eqref{eq:action} with respect to the dilaton field.
	So now we are ready to find the spacetime metric itself instead of the combination $B(z)$. 
	One can compute it by substituting the solution $\Phi(z)$ of \eqref{eq:reduced EE} into the definition of $B(z)$. The result for $\Phi_c=0$ is shown in FIG.~\ref{fig:metric}.
	% The metric is the solution of the Einstein-dilaton model as well as predicted from hadron spectra data.

	It is remarkable that the string frame metric has a valley. As is well known, such a metric profile gives a confinement potential of the holographic Wilson loop \cite{Maldacena:1998im,Rey:1998ik},
	and  we shall study the Wilson loop in the next section.
	On the other hand, the Einstein frame metric does not have such a minimum. 
	So the minimum is due to the dilaton field.
	Interestingly, this Einstein frame metric qualitatively matches that of the pure AdS metric.
	This result shows that the strange valley seen in the metric can be gravitationally consistent, as the phenomenon due to the frame difference.

		For any future use of our numerical results, we express the numerically obtained metric as a continuous function.
		Because we assume that the metric becomes the pure AdS spacetime at $z \to 0$, we fit the plot with a function including a $1/z^2$ term.
		As we compared our potential with IHQCD potential, we focus on the region $z<2$.
		One example of our fit results is 
		\\*
		\begin{align} \label{eq:fit gtt}
			g_{zz}(z) = \frac{1}{z^2} + 1.917 z^2.
		\end{align}
		\\*
		This appears to approximate well the valley around the bottom $z\sim 1$.

%%%%%%%%%%%%%%%%%%%%%%%%%%%%%%%%%%%%%%%%%%%%%%%

\section{Wilson loop}
	\label{sec:Wilson loop}

	Once the dilaton potential in the IHQCD action is determined, one can perform holographic calculation of some QCD quantities from the action. 
	In this section we calculate a holographic Wilson loop \cite{Maldacena:1998im,Rey:1998ik} for the bulk spacetime metric in FIG.~\ref{fig:metric}.

	Let us consider a static fundamental string which hangs down from the boundary of the spacetime and lies in the bulk.
	Defining $z_0$ as the deepest point of the string in the direction of $z$, the quark-antiquark potential $E$ and the quark-antiquark distance $d$ are computed by the following integrals:
	\begin{align}
		2\pi \alpha' E(z_0) &= 2 \int_{0}^{z_0} \!\! g_{zz}(z)\qty(\sqrt{\frac{g_{zz}(z)^2}{g_{zz}(z)^2 - g_{zz}(z_0)^2}} - 1) dz \nonumber \\
		\label{eq:quark potential}
		& ~~~~~~~~~~~~~~~~~~~~~ 
		- 2 \int_{z_0}^\infty \!\! g_{zz}(z) dz, \\
		\label{eq:quark distance}
		d(z_0) &= 2 \int_0^{z_0} \sqrt{\frac{g_{zz}(z_0)^2}{g_{zz}(z)^2 - g_{zz}(z_0)^2}}dz.
	\end{align}
	Here $\alpha'$ is the Regge slope parameter, and the gauge \eqref{eq:string frame metric} is assumed. After eliminating $z_0$ from these two integrals, one can finally obtain $E(d)$, the quark potential.

\begin{figure}[tb] 
		\centering
		\includegraphics[width=0.95\linewidth]{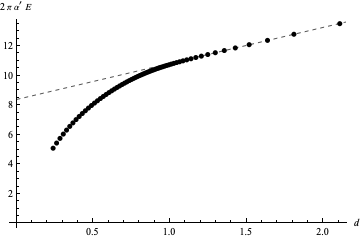}
		\caption{
			Quark-antiquark potential $2\pi\alpha'E(d)$ calculated with the string frame metric in FIG.~\ref{fig:metric}. Horizontal axis $d$ is measured in the unit of the AdS radius $L=0.51\text{[fm]}$\cite{Akutagawa:2020yeo}.
		}
		\label{fig:quark potential}
	\end{figure}

	The result for the determined holographic model FIG.~\ref{fig:metric}
	% the case $g_{zz}(z) = \exp (2A(z))$ 
	is shown in FIG.~\ref{fig:quark potential}.
	The linear behavior appears, as is expected from the shape of the string frame metric in FIG.~\ref{fig:metric}.
	The dashed line illustrates that the long-range part of calculated potential is almost linear.\footnote{
		To evaluate the gradient of this part is interesting because it is interpreted as a QCD string tension in the context of the sting flux tube model of hadrons, but this is out of the scope of this paper due to the unknown constant $\alpha'$.
	}
	We observe that the bulk reconstructed from the $\rho$ meson spectrum describes the quark confinement. 

\begin{figure}[tb]
		\centering
		\includegraphics[width=0.95\linewidth]{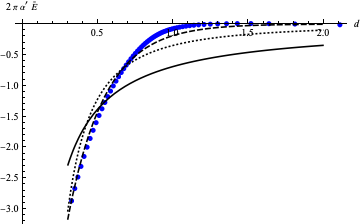}
		\caption{
			$E(d)$ is plotted with its linear contribution subtracted (blue points). We fit it by $\xi_1/d$ (solid line), $\xi_2/d^{\sigma_2}$ (dotted) and  $\xi_3 \exp(\sigma_3 d)$ (dashed) where $\xi_i$ and $\sigma_i$ are fitting parameters. 
		}
		\label{fig:qq_fit}
	\end{figure}

	Subsequently let us investigate the nonlinear part.
	To this end, we subtract the linear part from $E(d)$ and fit the remaining, defined as $\tilde{E}(d)$, by three kinds of simple functions.
	The best-fit results of each function are shown in FIG.\ref{fig:qq_fit} along with $\tilde{E}(d)$.
	% as seen in FIG.~\ref{fig:qq_fit}.
	We employ $1/d, 1/d^\sigma, \exp (\sigma d)$ as the model functions, among which the last one is found to be the best to approximate the nonlinear behavior.\footnote{
	The exponential suppression coincides with the result reported in the popular top-down models \cite{Kinar:1998vq} where other behavior was also studied.
	}

	Regarding the nonlinear part of the generic quark potential at large $d$, it is known that the leading correction following the linear part is order $1/d$, which is referred to as the L\"{u}scher term \cite{Luscher:1980ac}.
	% adding to linear potential from the holographic wilson loop by introducing the fluctuation of probe string and quantizing it\cite{Kinar:1999xu}.
	This behavior has been checked with the quark potential computed in lattice simulation \cite{Bali:1992ru}.
	\footnote{
		Early results in the comparison between the lattice result and the AdS/QCD model approach without quantum fluctuation are found in \cite{Andreev:2006ct,White:2007tu}.
	}
	In contrast, our result does not show the dependence of $1/d$ as seen in FIG.~\ref{fig:qq_fit}. 
	% shows the comparison between quark potential and some simple functions.
	This is still fine because the L\"{u}scher term is provided by a quantum fluctuation of the probe flux QCD string and thus of a fundamental string in the holographic perspective \cite{Kinar:1999xu} while we treat the probe string with no fluctuation so far.
	% As mentioned above, 
	In this sense L\"{u}scher term should not appear in our study, and our result is consistent with QCD.
	% Our data shows exponential behavior rather than that.

%%%%%%%%%%%%%%%%%%%%%%%%%%%%%%%%%%%%%%%%%%%%%%%

\section{Conclusion}
\label{sec:conclusion}

	In this paper we have derived a new dilaton potential for IHQCD based on the gravitational duals machine-learned from the meson spectrum.
	The Einstein equations \eqref{eq:EE} are 
	for the zero temperature metric \eqref{eq: string frame} and they include the dilaton potential $V(\Phi)$ to be determined. We have solved them by using the information of the machine-learned $B'(z)$ obtained in \cite{Akutagawa:2020yeo} as a constraint on the equations then have determined the explicit form of $V(\Phi)$.
	Its analytic form is well-fit by 
	a cosh function, 
	while interestingly the cosh form is
	the potential used to reproduce the equation of state of QCD in \cite{Gubser:2008ny}.
	Subsequently we have shown 
	that the dilaton potential satisfies the asymptotic condition \eqref{eq:IR behavior} with $(P,Q)=(1/2,2/3)$ which is 
	of the standard functional form for dilaton potentials employed
	in IHQCD.
	This asymptotic condition was determined phenomenologically for the glueball spectrum \cite{Gursoy:2007cb,Gursoy:2007er}, so it is surprising that the potential derived from the vector meson spectrum has turned out to satisfy the condition. Thus, as a sequence of processes, we 
	have been able to construct a reasonable IHQCD model in a data-driven way.

	Here we may make a comment on the fitting scheme of the obtained values of $(P,Q)$ in the derived dilaton potential.
	We have found 
	that $P$ and $Q$ depend on $\Phi_c$ in spite of the trivial dependence as shown in \eqref{eq:potential}. This happens because our fitting procedure only see a restricted region in $z$. The fit model \eqref{eq:IR behavior} includes powers in $\Phi$.   
		Then when the region of $\Phi$ for fit shifts by changing $\Phi_c$, the best fit parameters except for the overall constant also vary.
	This is a systematic issue to be improved, and would be resolved 
	if the bulk gravity can be more precisely reconstructed beyond $z=4$. 
	We may leave this as a future study.

	We would like to make another comment for those who have read the paper of the machine learning calculation \cite{Akutagawa:2020yeo}. 
	The AdS/QCD model assumed in this work can also be applied to higher spin meson \cite{,Katz:2005ir, Karch:2006pv}.
	For example one can apply it to spin 2 meson by considering the tensor field in the bulk.
	The rank of the tensor corresponds to the spin and $B'$ depends on the value of the spin.
	Therefore, by training the neural network also with the spectrum of the spin 2 meson, so called $a_2$ meson, the authors of \cite{Akutagawa:2020yeo} could determine the metric and dilaton individually.
	However our concern is that
	the results of the $a_2$ meson training did not capture the characteristics of the data well: 
	the training result
	could largely depends on the regularization imposed on the neural network.
	Thus in this work we have decided to use only the result of the training with the spectrum of $\rho$ meson, thus only a single $B'(z)$. Indeed, individual profiles of the dilaton field and the metric are not necessary for our procedure.
	Note that our result finds the confinement of the quark potential while it was not seen in the calculation incorporating the training results for the $a_2$ meson of \cite{Akutagawa:2020yeo}.

	Following the derivation of the IHQCD model, we have calculated 
	the Wilson loop, a QCD quantity, as a concrete prediction of the holographic model. 
	In this model, the metric has a structure with a valley in the bulk, where the probe string is expected to be stuck. 
	The conventional calculation of the holographic Wilson loop has lead to
	the quark-antiquark potential 
	linear in the long-range region. 
	For the nonlinear part of the potential, we have found that it 
	is approximated by an exponential function. 
	The origin of this behavior needs further study, but one explanation is by \cite{Hashimoto:2020mrx} in which it is shown that
	the leading nonlinearity of the holographic Wilson loop faster than $1/d$ always results
	in the valley, as is consistent with our emergent bulk metric in FIG.~\ref{fig:metric}.

	In this paper, we have treated only the Wilson loop as an observable 
	of the boundary theory, but IHQCD can make predictions for various thermodynamic quantities once the dilaton potential is given. 
	Thus by comparing our model with lattice QCD calculations, we may be able to find a better agreement on the thermodynamic quantities than the existing models. 
	We also have the advantage of being able to solve the equations again to derive a holographic model at finite temperature and calculate hydrodynamic observables describing the QGP phase. 
	Furthermore if a large number of various kinds of QCD data can derive a consistent dilaton potential, it leads to a construction of
	IHQCD models that describe QCD at various energy scales simultaneously.
	We will describe some models using different training QCD data in our future work \cite{ours2}, and hope that the combination of the holographic machine learning of quantum field theory data \cite{You:2017guh,Hashimoto:2018ftp,Hashimoto:2018bnb,Tan:2019czc,Hashimoto:2019bih,Hu:2019nea,Yan:2020wcd,Akutagawa:2020yeo,Hashimoto:2020jug,Song:2020agw} and further improvement of holographic QCD models will lift the final vail of ``the gravity dual of QCD."

\vspace*{20pt}

\begin{acknowledgments}
The authors are grateful to E.~Kiritsis, F.~Nitti and  for discussions.
	The work of K.~H.~is supported in part by JSPS KAKENHI Grant No.~JP17H06462.
	The work of T.~S.~is supported in part by JSPS KAKENHI Grant No.~JP20J20628.
\end{acknowledgments}

\bibliography{referrence}

\end{document}